\documentclass[twoside,leqno,twocolumn]{article}
\usepackage{ltexpprt}
\usepackage{times,amsmath,epsfig}
\usepackage{algorithmic}
\usepackage[ruled]{algorithm2e}
\usepackage{amssymb}
\usepackage{balance}

\begin{document}

\title{\Large On Graph Stream Clustering with Side Information}
\author{Yuchen Zhao\thanks{yzhao@cs.uic.edu, Department of Computer Science, University of Illinois at Chicago.} \\
\and
Philip S. Yu\thanks{psyu@uic.edu, Department of Computer Science, University of Illinois at Chicago; Computer Science Department,
King Abdulaziz University,
Jeddah, Saudi Arabia}}
\date{}

\maketitle

\newtheorem{definition}{Definition}


\begin{abstract}
Graph clustering becomes an important problem due to emerging applications involving the web, social networks and bio-informatics. Recently, many such applications generate data in the form of streams. Clustering massive, dynamic graph streams is significantly challenging because of the complex structures of graphs and computational difficulties of continuous data. Meanwhile, a large volume of side information is associated with graphs, which can be of various types. The examples include the properties of users in social network activities, the meta attributes associated with web click graph streams and the location information in mobile communication networks. Such attributes contain extremely useful information and has the potential to improve the clustering process, but are neglected by most recent graph stream mining techniques. In this paper, we define a unified distance measure on both link structures and side attributes for clustering. In addition, we propose a novel optimization framework {\em DMO}, which can dynamically optimize the distance metric and make it adapt to the newly received stream data. We further introduce a carefully designed statistics $SGS(C)$ which consume constant storage spaces with the progression of streams. We demonstrate that the statistics maintained are sufficient for the clustering process as well as the distance optimization and can be scalable to massive graphs with side attributes. We will present experiment results to show the advantages of the approach in graph stream clustering with both links and side information over the baselines.
\end{abstract}

\section{Introduction}
Recently, there is an increasing need for mining dynamic graphs with the rapidly growing social networks, Internet applications and communication networks \cite{DBLP:conf/sdm/AggarwalZY10}\cite{Kim:2009:PBE:1687627.1687698}\cite{DBLP:conf/sdm/Aggarwal11}\cite{Aggarwal:2011:ODG:2004686.2005654}\cite{Zhao:2011:GQE:2078331.2078335}. A graph stream is defined as individual graph objects arrive continuously over time, which represent various activities among nodes in the networks. Such activities can be discussion threads in social networks, user click graphs in user web browsing sessions and authorship graphs in a dynamically updated scientific repository.  The nodes of each graph object are typically drawn from a massive domain, such as users in social networks, IP addresses in Internet applications and terminals in communication networks. Although each graph object is in a modest size, the total number of distinct nodes and edges in the aggregated data from the stream can be extremely large.

\begin {figure} \center 
        \includegraphics[width=3in]{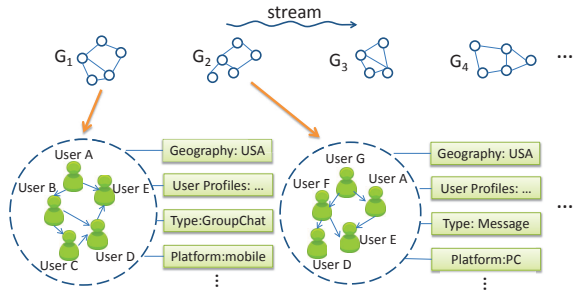} \\ 
    \caption{An Example of a (Directed) Social Activity Graph Stream with Side Information}
    \label{social_graph}
   \end {figure}

\begin {figure} \center 
        \includegraphics[width=3in]{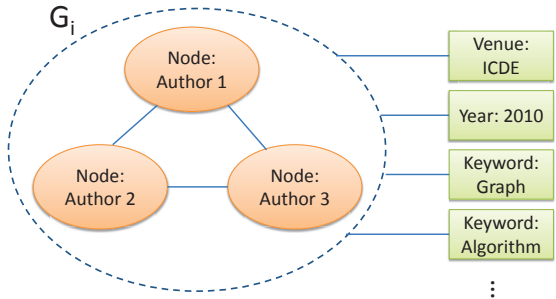} \\ 
    \caption{An Example of an (Undirected) Authorship Graph with Side Information}
    \label{dblp_graph}
   \end {figure}

Many existing approaches on graph stream mining have been devised to solve different tasks, including clustering \cite{DBLP:conf/sdm/AggarwalZY10}, classification \cite{DBLP:conf/sdm/Aggarwal11}, outlier detection \cite{Aggarwal:2011:ODG:2004686.2005654}, etc. All these approaches on graph streams are primarily designed for mining the link structures among graph objects only. However, in many real world applications, there are many side attributes associated with graphs that can be potentially highly useful to the mining tasks. Some examples of such attributes are listed as follows:
\begin{itemize}
\item In social networks, many social activities are generated daily in the form of streams, which can be naturally represented as graphs. In addition to the graph representation, there are tremendous side information associated with social activities, e.g. user profiles, behaviors, activity types and geographical information. These attributes can be quite informative to analyze the social graphs. We illustrate an example of such user interaction graph stream in Figure~\ref{social_graph}.
\item Web click events are graph object streams generated by users. Each graph object represents a series of web clicks by a specific user within a time frame. Besides the click graph object, the meta data of webpages, users' IP addresses and time spent on browsing can all provide insights to the subtle correlations of click graph objects.
\item In a large scientific repository (e.g. DBLP), each single article can be modeled as an authorship graph object \cite{DBLP:conf/sdm/AggarwalZY10}\cite{Aggarwal:2011:ODG:2004686.2005654}. In Figure~\ref{dblp_graph}, we illustrate an example of an authorship graph (paper) which consists of three authors (nodes) and a list of side information. For each article, the side attributes, including paper keywords, published venues and years, may be used to enhance the mining quality since they indicate tremendous meaningful relationships among authorship graphs.
\end{itemize}

Thus, it is much desired that the mining process can incorporate both links and side information to further improve its effectiveness. In this paper, we propose a framework to cluster graph streams with side information. The challenges of this problem are two-fold:

(1) Both graphs and side information are drawn from massive domains, which can not be explicitly held in the memory. For example, consider the number of users in a social network to be $N$. The potential number of distinct interactions (edges) can be as large as the order of $N^2$. Many side information such as IP address, tags, tokenized text and geographical information can also be extremely large. Even the summaries of those incoming data are rapidly growing with the streams and eventually unable to be explicitly stored. In addition, the problem becomes particular challenging in the stream scenario due to the high rate of incoming streams. Storing data in the hard disks and offline processing will not be able to efficiently handle the high volume of streams.

(2) Different types of side information give different indications on the nature of clustering, because many side attributes are quite noisy and insignificant. In other words, each side information type has its own degree that contributes to the underlining clustering. For example, while clustering individual graphs as shown in Figure~\ref{dblp_graph} from a large scientific repository, the aim is to group papers from the same research area into the same cluster. Thus, the links representing co-authorship as well as attributes of keywords and venues can be quite indicative to which cluster an individual graph object $G_i$ should be grouped into. However, the attributes including paper published years might not be that useful to cluster individual papers, since many research papers in different areas appear in various specialized conferences every year. Thus, considering both the linkage and side information, it is non-trivial to qualitatively measure the importance of each side information type as well as links.

In this paper, we first define a unified distance measure {\em E-S Distance} which combines the distances with regard to linkage and side information. Then we propose a novel optimization framework {\em DMO} which dynamically learns and tracks the importance (weights) of links and side attributes. The optimization framework on links and side attributes is periodically examined. It adjusts the weights to make the graphs within a cluster to be as coherent as possible while the graphs from different clusters to be as distinct as possible. Since efficiency is critical to stream algorithms and the data size is massive, it is not realistic to explicitly store all received data in the memory. We introduce a sketch based compression framework $SGS(C)$, which can store the statistics of {\em heterogenous data} including edges and side attributes. More importantly, we demonstrate that {\em DMO} can be efficiently and dynamically solved  in the sketch representation given by $SGS(C)$. We show that the proposed approach consumes constant memory with the growing incoming data and can be used to estimate all measures in the clustering algorithm as well as the optimization framework {\em DMO}.

The rest of this paper is organized as follows. In Section~\ref{related}, we discuss related work on graph stream clustering. In Section~\ref{dmo}, we define a unified distance metric {\em E-S Distance} on graph objects with side information. Then we present a novel optimization framework {\em DMO} to dynamically refine the distance measures with the progression of the stream. In Section~\ref{sketch}, we propose the statistics $SGS(C)$ and how to use $SGS(C)$ to estimate in the clustering algorithm. We report the experiment results in Section~\ref{experiment} and present the conclusion in Section~\ref{conclusion}.

\section{Related Work} \label{related}
In the literature, a number of techniques have been proposed to mine graph and network data \cite{Yan:2008:MSG:1376616.1376662}\cite{Kulis:2005:SGC:1102351.1102409}\cite{Dhillon:2005:FKM:1081870.1081948}\cite{Wang:2012:MCI:2339530.2339627}. Traditional graph clustering methods are extensively studied on the {\em node clustering} setting of a single static graph, including graph partitioning \cite{Kernighan70}, minimum cut \cite{Hao:1992:FAF:139404.139439}, heterogeneous networks \cite{Sun:2009:RCH:1557019.1557107}  and dense subgraph mining \cite{Zeng:2007:OCC:1242524.1242530}. The context of {\em node clustering} is to group similar nodes together based on linkage behaviors of a single large graph. Beside using only linkage information, Zhou et al. \cite{Zhou:2009:GCB:1687627.1687709} proposed a random walk based approach to cluster a single static graph by examining structural and attribute similarities. All these techniques are only applicable to the nodes in a {\em static individual} graph, rather than to cluster many graph objects whose nodes are drawn from a massive domain.

A number of approaches are also proposed in the context of {\em object clustering}, which are designed to cluster many graph objects.  The difference between {\em node clustering} and {\em object clustering} is that {\em object clustering} aims to cluster graph objects rather than nodes from a single graph. Many approaches have been proposed to discover the substructures of graphs \cite{Yan:2002:GGS:844380.844811}\cite{Zeng:2007:OCC:1242524.1242530}\cite{Zhao:2011:PUL:2117684.2118241}. However, mining subgraphs is computationally expensive with multiple passes. Therefore they are not applicable to handle continuous massive graph streams. \cite{Aggarwal:2007:XFP:1281192.1281201}\cite{dalamagas:clustering} are proposed to cluster objects in XML data.  However, those approaches cannot be scalable to a massive number of graphs and nodes, and are only able to handle disk-based data rather than stream data. Recently, a number of techniques are proposed to mine graph streams. \cite{DBLP:conf/sdm/AggarwalZY10} proposed a method to cluster massive graph streams by extending micro-clusters. \cite{DBLP:conf/sdm/Aggarwal11} is designed to construct the summary of graph streams and classify graph objects by scanning each of them only once. A structural connectivity model is proposed in \cite{Aggarwal:2011:ODG:2004686.2005654} to identify outliers in massive network streams. \cite{Zhao:2011:GQE:2078331.2078335} designed a graph sketch technique to estimate and optimize the queries on graph streams. However, all above approaches only consider the structure information among graphs, whereas neglect the side information associated with each graph object. Thus, many meaningful relationships and correlations of graph objects might not be discovered in the mining process.

Using side information to analyze record-based data within feature spaces are extensively studied in the context of {\em distance metric learning}. {\em Distance metric learning} studies the problem to learn proper distance metrics over inputs. \cite{Xing02distancemetric} proposed a global distance metric learning approach under a supervised setting. In addition, a number of approaches e.g. \cite{DBLP:conf/nips/DomeniconiG01}\cite{Peng02adaptivekernel} are designed to learn local adaptive distance metrics with supervised information. In an unsupervised setting, Principle Component Analysis (PCA) and Multiple Dimension Scaling (MDS) are widely used to reduce dimensions using a linear strategy. A detailed survey on {\em distance metric learning} can be found in \cite{yang2006distance}. However, these methods cannot be easily generalized to graph data, especially to dynamic graph streams.

\section{Distance Optimization} \label{dmo}

We first introduce some notations and definitions that will be used throughout the paper. Assume we have a stream of graphs ${\cal G}$ denoted as $\{G_1, G_2, ..., G_n, ...\}$, where each graph $G_i$ is drawn on the subset of massive nodes ${\cal N}$. We use set ${\cal E}$ to represent the set of distinct edges from all graphs: ${\cal E} = \{(X_1, Y_1), (X_2, Y_2), ..., (X_{n'}, Y_{n'}), ...\}$. Specifically, $X_j$ and $Y_j$ are the two nodes of each edge $(X_j, Y_j)$, and each graph $G_i$ contains a subset of edges from set ${\cal E}$. We assume the frequency of edge $(X_j, Y_j)$ in a graph $G_i$ is denoted by $F(X_j, Y_j, G_i)$. For example,
in communication networks, the frequency may represent the duration of conversations between two parties. The frequency may also be implicitly set to $1$ in many applications to reflect the link relationships between two nodes.

Associated with the graph stream, we also have $d$ different types of side information denoted by ${\cal T} = \{T_1, ..., T_d\}$. For example, the side information in authorship graph streams may contain different types of side attributes, such as publication years, conferences and paper keywords. We note that some attributes are associated with the whole graph while some attributes can also be associated with individual nodes or edges. We take the aggregated side attributes of nodes and edges, then append them to the whole graph. In addition, each graph may contain multiple side attributes of the same type, e.g. a paper has multiple keywords. Let ${\cal S}_l = \{S_{l1}, ..., S_{ln}, ...\}$ be all distinct side attributes of type $T_l$, where $l=1,...,d$. For example, if the type of $T_l$ is ``keyword", ${\cal S}_l$ stores all the distinct keywords appeared in the stream.  The value of side attribute $S_{ln}$ of type $T_l$ associated with graph $G_i$ is represented as $V(S_{ln}, G_i)$. For example, suppose the side attribute ``database" as a type ``keyword" appears in the paper $G_i$ for 3 times. Then its corresponding value is 3. Clearly, $V(S_{ln}, G_i)$ is 0 if graph $G_i$ does not contain the side attribute $S_{ln}$.

The goal of the stream clustering framework is to cluster graph objects into $k$ clusters, which are denoted by $C_1, C_2, ..., C_k$. Each incoming graph object from the stream is dynamically assigned to the most appropriate cluster, and the cluster is updated in real-time. Suppose we have a cluster $C_i$ containing a set of graphs $\{G_{i_1}, ..., G_{i_n}\}$. The implicit graph defined by the aggregation of graphs $\{G_{i_1}, ..., G_{i_n}\}$ is denoted by $H(C_i)$. In other words, $H(C_i)$ represents the summarization of graphs in cluster $C_i$. We use $N(C_i)$ to denote the number of graphs in cluster $C_i$.  The above notations are summarized in Table~\ref{notations}.

\begin{table}
\caption{Notations}
    \begin{tabular}{ | p{3.5cm} | p{4cm} |}
    \hline
    Symbol & Description \\ \hline
    ${\cal G}=\{G_1, ..., G_n, ...\}$ & graph streams \\ \hline
    ${\cal E}$ = $\{(X_1, Y_1)$ $, ...,$ $(X_{n'}, Y_{n'})$ $, ...\}$ & all distinct edges \\ \hline
    ${\cal T} = \{T_1, ..., T_d\}$ & $d$ types of side information with the stream\\ \hline
    ${\cal S}_l = \{S_{l1}, ..., S_{ln}, ...\}$ & all distinct side attributes of type $T_l$, where $l=1,...,d$ \\ \hline
    $F(X_j, Y_j, G_i)$ & the frequency of edge $(X_j, Y_j)$ \\ \hline
    $V(S_{ln}, G_i)$ & the value of side attribute $S_{ln}$ in type $T_l$ associated with $G_i$ \\ \hline
    $C_1, C_2, ..., C_k$ & $k$ clusters \\ \hline
    $H(C_i)$ & aggregated graphs in cluster $C_i$ \\ \hline
    $N(C_i)$ & the number of graphs in cluster $C_i$ \\ \hline
    \end{tabular} \label{notations}
\end{table}

\subsection{Preprocessing}
We propose a general framework to cluster both directed and undirected graphs. The edges can either be weighted or unweighted. The side information can also be of different formats. The notations used in the main paper implicitly assume each graph object in the stream is a directed graph. For undirected graphs, we convert them to directed graphs by applying lexicographic ordering on node labels. Thus, all notations can be simply reused for the case of undirected graphs after the conversion. In the meanwhile, we assign the frequencies of all edges to be $1$ if no frequency information is provided in the graphs.

We consider a general case that each side attribute is numeric. The binary and categorical attributes can be  converted to numeric attributes in a straightforward way. Specifically, binary attributes are special cases of numeric attributes. In addition, for categorical attributes, different categorical values can be assumed to be separate binary attributes. For side attributes that are associated with individual nodes or edges, we compute the aggregated values for each whole graph.

\subsection{Distance Definitions}
In order to cluster graph objects into a set of $k$ clusters such that similar graphs are grouped into the same clusters, a distance function is required to measure the similarities between graphs and clusters. Suppose we have a newly arrived graph $G_i$. For each cluster $C_j$ where $j = 1, 2, ..., k$, the distance between $G_i$ and $C_j$ is calculated by a distance function $d(G_i, C_j)$. The new graph $G_i$ will be grouped into its nearest cluster which has the minimum distance among all $k$ clusters. We note that each new graph contains both edge information and side information. Therefore, we define two types of distances for edge and side information respectively. The quadratic {\bf edge distance} between the graph $G_i$ and cluster $C_j$  is defined as:
\begin{eqnarray}
d^2_e(G_i, C_j)= \nonumber \\
\sum_{t=1}^m \left( F(X_t, Y_t, G_i) - \frac{F(X_t, Y_t, H(C_j))}{N(C_j)} \right)^2 \label{edge_dist}
\end{eqnarray}
where $m$ is the number of distinct edges received. In the above definition, all edges are enumerated and summed to the distance measure. However, we notice that only the edges contained in $G_i$ or $H(C_j)$ contribute to the summation. Since $H(C_j)$ is the summarization of all graphs in the $j$th cluster, the frequencies of edges $F(X_t, Y_t, H(C_j))$ are normalized by the number of graphs in the cluster. Similarly, the quadratic {\bf side distance} between the graph $G_i$ and cluster $C_j$ on the side information of type $T_l$ is defined as:
\begin{eqnarray}
d^2_s(G_i, C_j, T_l) = \nonumber \\
\sum_{t=1}^{m_l}
\left( V(S_{lt}, G_i) - \frac{V(S_{lt}, H(C_j))}{N(C_j)} \right)^2 \label{side_dist}
\end{eqnarray}
where $m_l$ is the number of distinct side attributes of type $T_l$.

Let vector $\vec{D}(G_i, C_j) = $ $[d_e(G_i, C_j)$, $d_s(G_i, C_j, T_1)$, $...$, $d_s(G_i, C_j, T_d)]^T$. The dimension of $\vec{D}(G_i, C_j)$ is $d+1$. Given the definitions of edge distance and side distances, we can further define the {\bf E-S distance} on \underline{e}dge and \underline{s}ide information:
\begin{definition} (E-S Distance)
The E-S distance on edge and side information is defined as:
\begin{eqnarray}
& & d^2(G_i, C_j) = \|G_i - C_j\|^2_A \nonumber \\
&=& w_0 \cdot d^2_e(G_i, C_j) + \sum_{l=1}^d \left( w_l \cdot d^2_s(G_i, C_j, T_l) \right) \nonumber \\
&=& \vec{D}(G_i, C_j)^T A   \vec{D}(G_i, C_j), \qquad A \succcurlyeq 0\label{combined}
\end{eqnarray}
where $w_0$ is the weight of edge distance and $w_l (l = 1, ..., d)$ are the weights of side distances. Matrix $A$ is a diagonal matrix $diag(w_0, w_1, ..., w_d)$ representing the weights of edge and side distances.
\end{definition}
In the above definition, matrix $A$ is required to be positive semi-definite $A \succcurlyeq 0$ to ensure the E-S distance be non-negative.

\subsection{Dynamic Multi-distance Optimization (DMO)}
A straightforward E-S distance measure may be using a $(d+1)$-dimensional identity matrix $I_{d+1} = diag(1, 1, ..., 1)$ as the matrix $A$. Thus, all edge distance and side distances are assigned to have equal weights $1$. However, different types of attributes and link information give different indications to the clustering. Suppose we cluster authorship graphs according to research areas. The coauthor relationships and paper keywords are clearly more important than author affiliations and publication years. The reasons are that researchers from the same area tend to collaborate, and papers of the same area are more likely to share the same keywords. Therefore, assigning equal or manually predefined weights to E-S distance cannot be generalized to vast real-world applications on massive graph streams.

In order to dynamically learn the weights of distances in matrix $A$ with the progression of the stream, we consider minimizing the {\bf intra-cluster distances} of graphs received so far:
\begin{eqnarray}
\operatorname*{min}_{A} \sum_{j=1}^k \sum_{G_i \in C_j}  \|G_i - C_j\|^2_A
\label{intra-obj-fun}
\end{eqnarray}
A trivial solution of this optimization problem is $A=0$. Thus, we further add a series of constraints to regulate the pairwise {\bf inter-cluster distances} between cluster centroids:
\begin{eqnarray}
\|C_i - C_j\|_A \geq c, \mbox{ for } i,j=1,...,k \mbox{ and } i \neq j \label{inter-constraint}
\end{eqnarray}
Here, the definition of inter-cluster distance $\|C_i - C_j\|_A$ is a natural extension of Eq.~\ref{combined}.\footnote{$L_2$-distance is not used here to prevent matrix $A$ always being rank 1.} $c$ in Eq.~\ref{inter-constraint} is an arbitrary positive constant which only affects the scales of weights. Thus we set $c$ to 1.  The optimization framework is given below:
\begin{eqnarray}
\operatorname*{min}_{A} & &   \sum_{j=1}^k \sum_{G_i \in C_j}  \|G_i - C_j\|^2_A \nonumber \\
s.t. & &
\|C_i - C_j\|_A \geq 1 \qquad (i,j=1,...,k \mbox{ and } i \neq j) \nonumber \\
& & A \mbox{ is diagonal}
, \qquad A \succcurlyeq 0
\label{optimization}
\end{eqnarray}
The idea of the optimization is to let the graphs within the same clusters to be as coherent as possible and graphs from different clusters can be separated well.

\begin{lemma} \label{lemma_convex}
The proposed optimization framework in Eq.~\ref{optimization} is a convex optimization problem.
\end{lemma}
\begin{proof}
From the definition in Eq.~\ref{combined}, it is clear that the objective function is a linear function on $A$. Thus the objective function is a convex function. The inter-cluster distance constraints can be rewritten as $1 -  \left( \|C_i - C_j\|_A^2 \right)^{1/2} \leq 0$ for $i,j=1,...,k \mbox{ and } i \neq j$. Since $\left( \|C_i - C_j\|_A^2 \right)^{1/2}$ is concave, the inter-cluster distance constraints are convex functions. It is also straightforward to verify the constraints on matrix $A$ are convex \cite{Boyd:2004:CO:993483}. Thus, the optimization framework in Eq.~\ref{optimization} is a convex optimization problem.
\end{proof}

We propose {\bf D}ynamic {\bf M}ulti-distance {\bf O}ptimization {\bf (DMO)} to solve the above optimization framework, and we use DMO to dynamically refine the weights of graph edges and side attributes.

\begin{lemma} [{\bf DMO}] \label{lemma_dmo}
The solution of proposed optimization framework in Eq.~\ref{optimization} can be approximated by solving the following form:
\begin{eqnarray}
& & \operatorname*{min}_{A}  \qquad t \sum_{j=1}^k \sum_{G_i \in C_j}  \|G_i - C_j\|^2_A - \nonumber \\
& & \sum_{i=1}^k \sum_{j=1, j \neq i} log \left(   \|C_i - C_j\|_A - 1 \right),  t > 0, A \succcurlyeq 0 \label{log-barrier}
\end{eqnarray}
\end{lemma}
\begin{proof}
Since the inter-cluster distance inequality constraints can be rewritten as $1 -  \left( \|C_i - C_j\|_A^2 \right)^{1/2} \leq 0$, we define the log-barrier of the problem as:
\begin{eqnarray}
\phi(A) = - \sum_{i=1}^k \sum_{j=1, j \neq i} log \left(   \|C_i - C_j\|_A - 1 \right)
\end{eqnarray}
Eq.~\ref{log-barrier} can be directly derived by applying the log-barrier to the objective function in Eq.~\ref{optimization} \cite{Boyd:2004:CO:993483}. Here, $t$ is a positive parameter of the logarithmic barrier method.
\end{proof}

We will describe the details on efficient distance estimation in Section \ref{sketch}. Suppose the distances on edges and side information are available, Eq.~\ref{log-barrier} in DMO can be solved by using the gradient descent algorithm. Specifically, the matrix $A$ is initialized to be an identity matrix which gives all distances equal weights. For every $\gamma$ graphs clustered, a gradient descent search is applied to Eq.~\ref{log-barrier} and weights are dynamically optimized based on the newly received graph edges and side attributes. By enabling DMO, the adjusted weights ensure that the intra-cluster distances are minimized and the inter-cluster distances are maximized. Thus, the weights of both edge and side information can be gradually and dynamically refined throughout the streams.\footnote{Since the weights are refined gradually, every update takes only a few search steps. We do not use the Newton method because it is usually slower due to the matrix inverse at each update.}

\section{Sketch-Based Clustering Framework} \label{sketch}
One challenge of stream mining is the growing size of available data. This problem is especially critical on the graph data with side information. On the one hand, graphs are drawn from a massive set of nodes in many real applications, and the number of possible edges are quadratic with the number of nodes. On the other hand, the volume of side information can also be quite large. Furthermore, both the sizes of edge and side information are growing with more and more data received. When the sizes become extremely large, this brings enormous difficulties to maintain all data in the memory. In this section, we propose a carefully designed sketch-based framework to maintain the statistics of incoming data. The proposed framework considerably reduces the storage requirement and only requires constant memory spaces with the streams.
We also demonstrate how to use the statistics maintained to accurately estimate the key measures in the clustering process as well as the optimization framework {\em DMO}.

\subsection{Preliminaries}
Sketch approaches are generic methods to approximate aggregation functions in the data stream domain. We adapt Count-Min sketch \cite{Cormode:2005:IDS:1073713.1073718} to estimate frequency statistics of data points, and extend it to the context of graphs with side information. 
Sketch approaches are generic methods to approximate aggregation functions in the data stream domain. We adapt Count-Min sketch \cite{Cormode:2005:IDS:1073713.1073718} to estimate frequency statistics of data points, and extend it to the context of graphs with side information. In each sketch table, we maintain a two-dimensional array with $w \cdot h$ cells with $w= \lceil \mbox{ln}(1/\delta) \rceil$ rows and $h=\lceil e/\epsilon \rceil$ columns, where $e$ is the base of the natural logarithm. In addition, $w$ different hash functions $f_1, ..., f_{w}$ are randomly generated from a pairwise-independent family. Each hash function corresponds to one of 1-dimensional row arrays with $h$ cells in the sketch. When a new data point $d_i$ arrives, each hash function $f_j$ is applied to $d_i$ and maps it to a hash value $v_j$ with range $[0, h-1]$. For the $j$th hash function, the frequency of data point $d_i$ is added to the $v_j$th column on the $j$th row of the sketch. Thus, only one cell on each row is updated, and there are $w$ cells in the sketch table that are incremented by the frequency of $d_i$.

In order to estimate the frequency of a data point, we map the data point to $w$ cells in the sketch table by applying the $w$ hash functions. The frequency of the data point is determined by the minimum value among all these $w$ cells. We notice that the sketch table can only overestimate the actual values, since the frequencies are non-negative and cells are updated by addition. As shown in \cite{Cormode:2005:IDS:1073713.1073718}, the estimate guarantees that the overestimate is no more than $\epsilon \cdot T$ with probability at least $1-\delta$ for a data stream with $T$ arrivals. This {\em probabilistic upper bound} shows that increasing $w$ and $h$ can get more accurate estimation. Its sensitivity on the size of the sketch table has been studied in previous work \cite{Aggarwal:2011:ODG:2004686.2005654}\cite{DBLP:conf/icde/AggarwalZY12}. In the following, we will present how to apply sketches to statistics maintenance on graph streams with side information.

\subsection{Sketch Based Statistics}
Instead of storing the explicit edges and side information, we maintain the following statistics:
\begin{definition} \label{stats}
The {\bf S}tatistics of {\bf G}raphs with {\bf S}ide information {\bf SGS(C)} maintained in the memory  for each cluster $C$ is defined as \{$ESketch(C)$, $ER(C)$, $SSketch(C, 1...d)$, $SR(C, 1...d)$, $N(C)$, $T(C)$\}. Each component in  SGS(C) is defined in details as:
\begin{itemize}
\item $ESketch(C)$. one $w \cdot h$ sketch table storing first moments of edge frequencies.
\item $ER(C)$. the summation of second moments of edge frequencies: $ER(C) = \sum_{G_i \in C} \sum_{t=1}^m F^2(X_t, Y_t, G_i)$.
\item $SSketch(C, 1...d)$. $d$ $w \cdot h$ sketch tables storing first moment values for $d$ side attribute types correspondingly.
\item $SR(C, 1...d)$. a vector with length $d$ containing the summation of second moments of side attribute values: $SR(C, l)$ $=$ $\sum_{G_i \in C} \sum_{t=1}^{m_l} V^2(S_{lt}, G_i)$, $l = 1,...,d$.
\item $N(C)$. the number of graphs in the cluster $C$.
\item $T(C)$. the most recent timestamp of the cluster being updated.
\end{itemize}
\end{definition}

When a new incoming graph $G_t$ is assigned to a cluster $C$, the statistics in $SGS(C)$ are updated as follows. For each edge $(X_i, Y_i)$, $w$ hash functions are applied to $X_i \oplus Y_i$ and the hash values are used to determine $w$ cells in the sketch table $ESketch(C)$. Here, $\oplus$ is the concatenation operator on the node label strings. Those $w$ cells are incremented by $F(X_i, Y_i, G_t)$. In the meanwhile, the second moment of its edge frequency is added to $ER(C)$. Similarly, each side attribute in graph $G_t$ is hashed into $SSktch(C, 1...d)$, and $SR(C, 1...d)$ is updated based on the second moment value. Lastly $N(C)$ is incremented by $1$ and $T(C)$ is updated to the current time.

Since none of the components' sizes in $SGS(C)$ grow in the update, the statistics maintained always keep a {\em constant} storage with the progression of the stream. Furthermore, another advantage of $SGS(C)$ is that the storage used by $SGS(C)$ can be easily adjusted by setting the sizes of sketches to adapt the local hardware requirement.
We further observe that $SGS(C)$ follows the additive property:
\begin{lemma} \label{lemma_stats}
The statistics maintained in Definition~\ref{stats} follows the additive property. In other words, $SGS(C_1 \cup C_2)$ can be computed as a function of $SGS(C_1)$ and $SGS(C_2)$.
\end{lemma}
\begin{proof}
The sketch table in $ESketch(C_1 \cup C_2)$ can be computed by additions of two-dimensional arrays in $ESketch(C_1)$ and $ESketch(C_2)$. Similarly, $SSketch(C_1 \cup C_2, 1...d)$ is also the summation of sketch tables $SSketch(C_1, 1...d)$ and $SSketch(C_2, 1...d)$. For second moments,
\begin{eqnarray}
&&ER(C_1  \cup C_2) \nonumber \\
&=& \sum_{G_i \in (C_1  \cup C_2)} \sum_{t=1}^m F^2(X_t, Y_t, G_i) \nonumber \\
&=& \sum_{G_i \in C_1} \sum_{t=1}^m F^2(X_t, Y_t, G_i) + \sum_{G_i \in C_2} \sum_{t=1}^m F^2(X_t, Y_t, G_i)\nonumber \\
&=& ER(C_1) + ER(C_2) \nonumber
\end{eqnarray}
Likewise, $SR(C_1  \cup C_2, 1...d) = SR(C_1, 1...d) + SR(C_2, 1...d)$. $N(C_1 \cup C_2)$ is the number of total graphs in $C_1$ and $C_2$, thus $N(C_1 \cup C_2) = N(C_1) + N(C_2)$. $T(C_1 \cup C_2)$ is the most recent timestamp of $C_1$ and $C_2$, hence $T(C_1 \cup C_2) = max (T(C_1), T(C_2))$ .
\end{proof}

\subsection{Algorithm with Side Information (GSSClu)}
Here, we present the algorithm for clustering graph streams with side information. The input of the clustering algorithm is the number of clusters $k$. The only information we store is the set of cluster statistics $\{SGS(C_1), ..., SGS(C_k)\}$. As the initialization step, we set $A$ to be an identity matrix which gives equal weights to the edge and side information. For the first $k$ received graphs, we create $k$ singleton cluster statistics $SGS(C_i), i=1,...,k$ respectively. While the initialization may not create a well-separated clustering, these $k$ clusters will be further stabilized in the subsequent steps.  For each new graph $G_i$, we compute the {\em E-S distance} on edge and side information between $G_i$ and $k$ clusters. Assume $C_{min}$ is the closest cluster to $G_i$ among all $k$ clusters. We also want to measure the {\em structural spread} of the cluster $C_{min}$ since $G_i$ may not necessarily belong to cluster $C_{min}$. The reason is that $G_i$ might be an outlier or represent a new cluster, despite that it has the shortest distance to $C_{min}$ compared with other clusters. Thus, we define the {\bf structural spread} of a given cluster $C_j$ as a function of the mean square radius of $C_j$:
\begin{eqnarray}
S(C_j) = \frac{p}{N(C_j)} \sum_{G_i \in C_j}  \|G_i - C_j\|^2_A
\label{spread}
\end{eqnarray}
Here, the spread $S(C_j)$ is defined as the mean square radius of the cluster $C_j$ multiplied with a factor $p$.\footnote{We use $p=3$ in accordance with the normal distribution assumption.}
If the graph $G_i$ is within the spread of $C_{min}$, $G_i$ is assigned to cluster $C_{min}$ and the statistics of $SGS(C_{min})$ is updated accordingly. Otherwise, the graph $G_i$ may be an outlier or represent a new born cluster. Therefore, we remove the most stale, namely least recently updated, cluster based on the stored timestamps, and create singleton cluster statistics from $G_i$. In the meanwhile, for every $\gamma$ graphs obtained from the stream, we dynamically optimize the matrix $A$ based on newly received information using Eq.~\ref{log-barrier} defined in {\em DMO}. Thus, the weights of both edge and side information can be actively learned and adjusted with the evolving stream. The detailed description on the clustering method {\em GSSClu} can be found in Algorithm~\ref{alg:one}.

\begin{algorithm}[t]
\KwIn{$k$: number of clusters}
Initialize  cluster statistics set to be an empty set; \\
$A = I_{d+1}$;\\
$graph\_count = 0$;\\
\ForEach{newly received graph $G_i$}{
    $graph\_count = graph\_count + 1$;\\
    \If{$i < k$}{
        create singleton cluster statistics $SGS(C_i)$ by inserting $G_i$; \\
        continue;
    }
    \For{$j = 1$ to $k$}{
        compute $\|G_i - C_j\|^2_A$ defined in Eq.~\ref{combined};
    }
    let $C_{min}$ be the closest cluster; \\
    \lIf{$\|G_i - C_{min}\|^2_A < $ $S(C_{min})$}{
        assign $G_i$ to $SGS(C_{min})$
    }\;
    \lElse{
        replace least recently updated cluster statistics
        by singleton cluster statistics created from $G_i$;
    }

    \If{$graph\_count$ $\%$ $\gamma == 0$}{
    adjust $A$ by optimizing Eq.~\ref{log-barrier};
    }
}
\caption{Clustering Graph Streams with Side Information (GSSClu)}
\label{alg:one}
\end{algorithm}

\subsection{Key Measures Estimation}
Next, we will illustrate how to compute the measures in {\em GSSClu} using the statistics maintained by $SGS(C)$.

\begin{lemma} \label{lemma_sgc}
The statistics maintained in $SGS(C_j), j=1,...,k$ are sufficient to compute all measures required by the clustering algorithm {\em GSSClu}.
\end{lemma}

\begin{proof}
From Algorithm~\ref{alg:one}, it is clear that the clustering process requires the following measures:
\begin{itemize}
\item $\|G_i - C_j\|^2_A$ (Eq.~\ref{combined}): the E-S distance between a newly received graph $G_i$ and a cluster $C_j$.
\item $\sum_{G_i \in C_j}  \|G_i - C_j\|^2_A$ (Eq.~\ref{intra-obj-fun}): the intra-cluster distance of a cluster $C_j$ where $G_i$ represents all graphs clustered in $C_j$.
\item $\|C_i - C_j\|_A$ (Eq.~\ref{inter-constraint}): the inter-cluster distance between two clusters.
\item $S(C_j)$ (Eq.~\ref{spread}): the structural spread of a cluster $C_j$.
\end{itemize}
All these four distance measures are computed based on combinations of side information distances and edge distance. We will only show how to compute the measure related to the side information due to the space limitation. The computation in terms of the edges can be derived in a similar way.

{\bf E-S Distance of New Graphs:} The distance between an {\em incoming} graph $G_i$ and a cluster $C_j$ on the side information of type $T_l$ is defined in Eq.~\ref{side_dist}. It can be expanded as:
\begin{eqnarray}
& & d^2_s(G_i, C_j, T_l) \nonumber \\
&=& \sum_{t=1}^{m_l} \left( V(S_{lt}, G_i) - \frac{V(S_{lt}, H(C_j))}{N(C_j)} \right)^2 \nonumber \\
&=& \sum_{t=1}^{m_l}  V^2(S_{lt}, G_i) - \sum_{t=1}^{m_l} 2 V(S_{lt}, G_i)\frac{V(S_{lt}, H(C_j))}{N(C_j)} \nonumber \\
& & + \sum_{t=1}^{m_l} \frac{V^2(S_{lt}, H(C_j))}{N^2(C_j)}  \label{side_dist_exp}
\end{eqnarray}
Since $G_i$ is the newly received graph from the stream, its side attributes are available and known exactly. Therefore, the first term $V^2(S_{lt}, G_i)$ in Eq.~\ref{side_dist_exp} can be computed exactly. In the second term, only a non-zero value of both $V(S_lt, G_i)$ and $V(S_lt, H(C_j))$ will add up to the summation. Hence, instead of computing all $m_l$ side attributes, we only need to enumerate all side attributes contained in $G_i$. $V(S_lt, H(C_j))$ can be directly estimated from the sketch table $SSketch(C_j, l)$ in $SGS(C_j)$, whereas the exact value of $V(S_{lt}, G_i)$ is known. $N(C_j)$ is also stored in $SGS(C_j)$. The third term can be computed by performing pairwise self products of each row in $SSketch(C_j, l)$. The minimum of these $w$ values divided by $N^2(C_j)$ is used as the estimate value.


{\bf Intra-cluster Distance:}
The intra-cluster distance of cluster $C_j$ is defined as the sum of distances between every graph clustered in $C_j$ and $C_j$'s centroid. Different from the previous computation in Eq.~\ref{side_dist_exp}, one should note that the graphs clustered in $C_j$ are not explicitly stored. Hence, the estimation from Eq.~\ref{side_dist_exp} cannot be directly reused. For the side information of type $T_l$, it can be expanded as:
\begin{eqnarray}
& & \sum_{G_i \in C_j} d^2_s(G_i, C_j, T_l) \nonumber \\
&=& \sum_{G_i \in C_j} \sum_{t=1}^{m_l} \left( V(S_{lt}, G_i) - \frac{V(S_{lt}, H(C_j))}{N(C_j)} \right)^2 \nonumber \\
&=& \sum_{G_i \in C_j} \sum_{t=1}^{m_l} V^2(S_{lt}, G_i) - 2 \sum_{t=1}^{m_l} \frac{V^2(S_{lt}, H(C_j))}{N(C_j)} \nonumber \\
& & + N(C_j) \sum_{t=1}^{m_l} \frac{V^2(S_{lt}, H(C_j))}{N^2(C_j)} \nonumber \\
&=& SR(C_j, l) - \frac{\sum_{t=1}^{m_l}  V^2(S_{lt}, H(C_j))}{N(C_j)} \label{intra_dist}
\end{eqnarray}
From Eq.~\ref{intra_dist}, one can observe that $SR(C_j, l)$ and $N(C_j)$  are both explicitly maintained in $SGS(C_j)$. $\sum_{t=1}^{m_l}  V^2(S_{lt}, H(C_j))$ in the second term can be estimated by pairwise products of each row in $SSketch(C_j, l)$. Thus, the intra-cluster distance can also be computed from the statistics maintained.

{\bf Inter-cluster Distance:}
The inter-cluster distance is defined as the distance between two clusters' centroids in Eq.~\ref{inter-constraint}. The inter-cluster distance in terms of side information $T_l$ is:
\begin{eqnarray}
& & d^2_s(C_i, C_j, T_l) \nonumber \\
&=& \sum_{t=1}^{m_l} \left( \frac{V(S_{lt}, H(C_i))}{N(C_i)} - \frac{V(S_{lt}, H(C_j))}{N(C_j)} \right)^2 \nonumber \\
&=& \sum_{t=1}^{m_l}  \frac{V^2(S_{lt}, H(C_i))}{N^2(C_i)} - \sum_{t=1}^{m_l} 2 \frac{V(S_{lt}, H(C_i)) V(S_{lt}, H(C_j))}{N(C_i) N(C_j)} \nonumber \\
& & + \sum_{t=1}^{m_l} \frac{V^2(S_{lt}, H(C_j))}{N^2(C_j)}  \label{inter_dist_exp}
\end{eqnarray}
As shown previously, the first and third terms can be computed by pairwise self products of sketches in $SGS(C_i)$ and $SGS(C_j)$ respectively. Similarly, the second term can be computed by the product of each row from $SSketch(C_i,l)$ and $SSketch(C_j,l)$, and the minimum of $w$ rows is used as the estimate value.

{\bf Cluster Structural Spread:}
From the definition in Eq.~\ref{spread}, the spread of a cluster $C_j$ related to the side information type $T_l$ can be represented as:
\begin{eqnarray}
\frac{p}{N(C_j)} \sum_{G_i \in C_j} d^2_s(G_i, C_j, T_l)
\end{eqnarray}
Since $\sum_{G_i \in C_j} d^2_s(G_i, C_j, T_l)$ can be estimated from Eq.~\ref{intra_dist}, the structural spread can be also computed from the statistics.

Therefore, all measures used in Algorithm~\ref{alg:one} including {\em DMO} can be estimated by the statistics maintained in $SGS(C_j), j = 1,...,k$. We further observe that the accuracies of estimations are directly related to the sketches, which are bounded by the {\em probabilistic upper bound} described earlier. 
\end{proof}

\section{Experimental Results} \label{experiment}
In this section, we present the effectiveness and efficiency of the proposed clustering scheme with a number of baselines on real data sets. We refer to our approach as the {\em GSSClu} method, since it is designed for \underline{G}raph \underline{S}tream  with \underline{S}ide Information \underline{Clu}stering.

\subsection{Data Sets}
We use two real data sets, namely CORA and IMDB, to evaluate the {\em GSSClu} method. We use two real data sets to evaluate the {\em GSSClu} method. The details of these two data sets are listed as follows:

\begin{itemize}
 
\item {\bf CORA Data Set:} The first data set that we use in the evaluation is the CORA data set\footnote{http://www.cs.umass.edu/\textasciitilde{}mccallum/code-data.html}. The CORA data set consists of 19,396 scientific articles in the computer science domain. In order to compose author-pair graph streams from the scientific publications, we consider each scientific article as a graph object with co-author relationships as edges as in \cite{DBLP:conf/sdm/AggarwalZY10}\cite{Aggarwal:2011:ODG:2004686.2005654}. We use the research topics of research papers as the ground truth to evaluate the clustering quality. In the CORA data set, all research papers are classified into a topic hierarchy, with 73 sub topics on the leaf level. We use the second level topics as the labels to evaluate. There are 10 topics in total, which are  {\em Information Retrieval, Databases, Artificial Intelligence, Encryption and Compression, Operating Systems, Networking, Hardware and Architecture, Data Structures Algorithms and Theory, Programming} and {\em Human Computer Interaction}. Each paper has an average 3.3 authors. For the side attributes, we obtain two types of side information to assist clustering: terms and citations. The terms are extracted from the paper titles, and citations include a list of papers that a given article cites. One paper cites 4.3 papers and has 6.1 distinct terms in average.

\item {\bf IMDB Data Set:} The {\bf I}nternet {\bf M}ovie {\bf D}ata{\bf b}ase is an online collection of movies and television shows, which also contains the related information, such as actors, directors, production crew, etc. We obtain a sample of IMDB data set, which covers ten-year movie data from the year of 1996 to 2005 in USA. We further process the data set to compose an actor-pair graph stream from it. Each movie is considered as a graph object. Actors of the movie are nodes and actor-pairs are considered as edges within the graph. In order to evaluate the effectiveness of the proposed clustering method, we use the movie genre as the label. We extract the movies of the top four genres from the IMDB, including {\em Short}, {\em Drama}, {\em Comedy} and {\em Documentary}. In addition, we remove the movies which have more than one label. In total, there are 9,793 movies, which consist of 1,718 movies from the {\em Short} genre, 3,359 movies from the {\em Drama} genre, 2,324 movies from the {\em Comedy} genre and 2,392 movies from the {\em Documentary} genre. One movie graph object has 24.6 edges in average.

Moreover, we extract three side information types associated with the actor-pair graphs, namely plot keywords, producers and directors. We extract words by tokenizing movie plots. After stop words removal, the frequent words are used as keywords. We notice that the distinct keywords, producers and directors from the whole data set are very large due to the data sparsity. In average, one movie graph object has 16.3 keywords, 1.1 producers and 1.4 directors.
\end{itemize}

\subsection{Methods}
In order to demonstrate the effectiveness and efficiency of the proposed approach, we compare {\em GSSClu} with a number of baselines. Since there is no known method to cluster graph streams with side information, we use the following approaches to show the performance of {\em GSSClu} from different perspectives:

(1) {\bf GMciro:} \cite{DBLP:conf/sdm/AggarwalZY10} proposed {\em GMicro} to cluster graph streams by extending the micro-cluster model. {\em GMicro} is the best known method to cluster fast and high volume graph streams by considering the similarities of edge structures. However, this method only considers the linkage within graphs, and cannot utilize the massive side information associated with graphs to enhance the clustering process.
 
(2) {\bf GSSClu [w/o opt.]:} Since {\em GMicro} does not use side information to cluster, we use a variation of {\em GSSClu} to demonstrate the power of dynamic distance optimization framework {\em DMO} as shown in Eq.~\ref{log-barrier} for a fair comparison. Instead of dynamically optimizing the importance among links and various side attributes, this method assigns them with equal weights as a simplified version of {\em GSSClu}. We refer to this approach as {\em GSSClu [w/o opt.]} in all following figure legends.
 
(3) {\bf Disk-based GSSClu:} In order to show the effectiveness of the proposed sketch-based framework {\em SGS(C)}, we develop another variation of {\em GSSClu} by computing the exact values of all metrics in the clustering algorithm. Due to the massive size of the incoming stream data and its growing nature, the data can only be stored on the hard disk to avoid the out of memory problem. We refer to this approach as {\em Disk-based GSSClu}. By comparing it with {\em GSSClu}, we can understand how close the sketch-based framework {\em SGS(C)} can estimate the true values. However, one should note that {\em Disk-based GSSClu} is about 5 to 10 times slower than {\em GSSClu} due to the long response time of disk queries.

\subsection{Metrics and Settings}
The goal of the evaluation is to examine if the proposed approach can effectively use linkage and side information from the streams to improve the clustering results over the baselines. In order to test the effectiveness of the proposed scheme, we use the {\em cluster purity measure} \cite{DBLP:conf/sdm/AggarwalZY10} to evaluate the clustering quality. For each data set, the labels of graph objects are known but excluded from the clustering process. We only use the labels to measure the quality of clustering. Specifically, for each generated cluster, we compute the dominate class labels from the graph objects within the cluster. The purity of each cluster is computed as the fraction of graph objects in the cluster which belong to the dominate class label. We report the average purity scores of different clusters as the {\em cluster purity measure}. We note that the cluster purity ranges from 0 to 1, and 1 represents a perfect clustering result. Clearly, a good clustering will provide a high value of the cluster purity. For efficiency, we report the processing rate for the proposed method and baselines.

Unless otherwise mentioned, the default parameter $\gamma$ is set to be 250. We also test the sensitivity over $\gamma$ at the end of this section. The number of hash function is set to be 10 and the width of sketch table is set to be 500 for all baselines and the proposed method. The default values of $k$ are 10 for the CORA data set and 8 for the IMDB data set, which are the same settings as in \cite{Aggarwal:2011:ODG:2004686.2005654}\cite{DBLP:conf/icde/AggarwalZY12}.

\subsection{Effectiveness Results}
   \begin {figure}  \vspace{-0.4in}
      \begin{tabular}{cc}
        \includegraphics[width=1.6in]{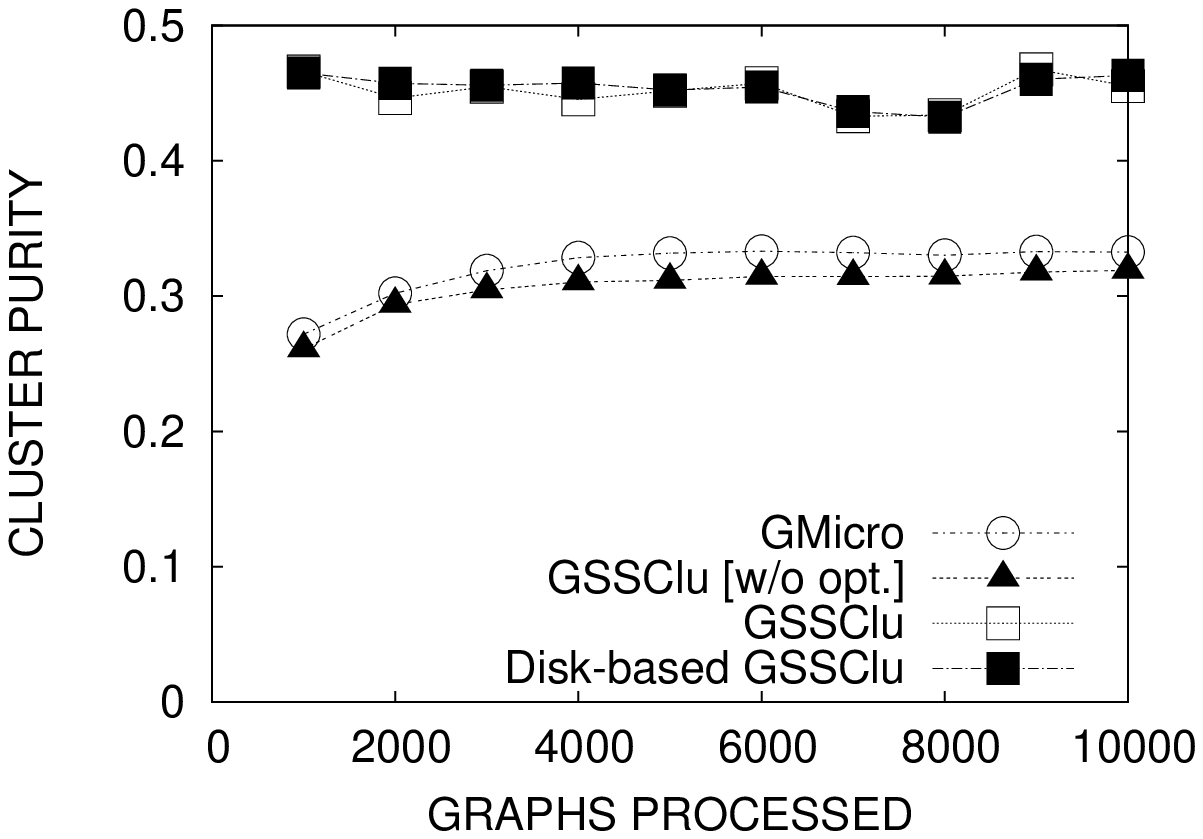} &  \includegraphics[width=1.6in]{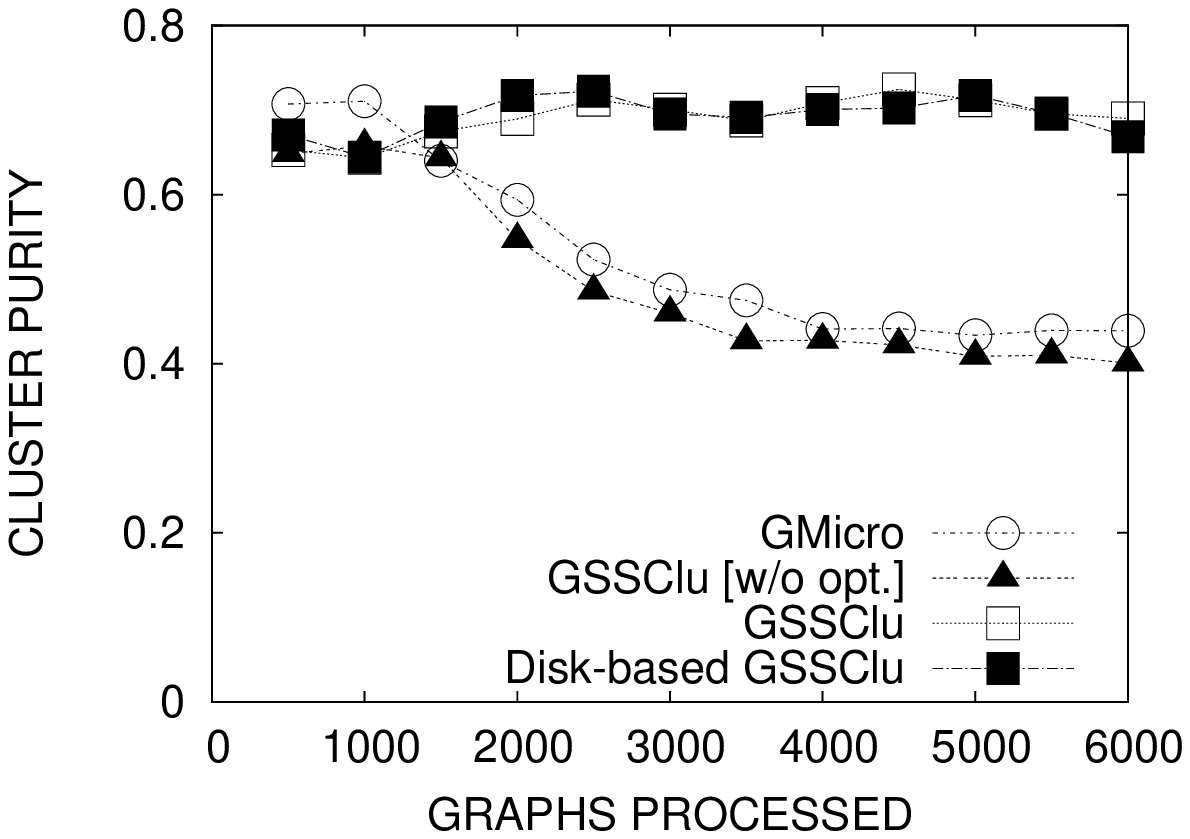} \\
      (a) CORA Data Set & (b) IMDB Data Set \\
      \end{tabular}
    \caption{Cluster Purity}\label{purity}
   \end {figure}
We first show the effectiveness results for the CORA and IMDB data sets. The effectiveness results for {\em GSSClu} and baseline algorithms with increasing number of processed graphs are shown in Figures~\ref{purity} (a) and (b). The number of graphs processed is shown on the X-axis, whereas the cluster purity is illustrated on the Y-axis.

For the CORA data set, we can observe that all four approaches achieve stable performance along the progression of the stream. The reason is that all these four approaches are designed to process stream data. Since the CORA data set has 10 labels, a random assignment will generate clusters with purity roughly at 0.1. From the figure, {\em GMicro} achieves about 0.33 cluster purity by using links only. The performance of {\em GSSClu [w/o opt.]} is lower than {\em GMicro} although it uses both the links and side information. This is because side information sometimes are quite noisy. Thus, assigning arbitrary (in this case, equal) weights to links and side attributes may even degrade the clustering quality. We further note that the proposed approach {\em GSSClu} has a purity score at around 0.45, which gains a performance at least 10\% over {\em GMicro} and {GSSClu [w/o opt.] in terms of purity. This suggests that the distance optimization {\em DMO} can indeed effectively learn the importance among links and different side attributes. In the meantime, the performances of {\em GSSClu} and {\em Disk-based GSSClu} are quite similar. {\em Disk-based GSSClu} is only slightly higher than {\em GSSClu} in term of purity, which can hardly be distinguished from the figure. This further demonstrates that the sketch-based approximation maintains the accuracy of the clustering process.

For the IMDB data set, the four approaches get similar performances for the first 1,000 received graphs. {\em GMicro} gives an even higher purity score than {\em GSSClu}. The reason of this is that {\em GSSClu} does not get enough statistics to infer the importance of links and side attributes with limited data. With more and more graphs received, we can observe that {\em GSSClu} significantly outperforms both {\em GMicro} and {\em GSSClu [w/o opt.]} with at least 0.25 purity improvement. We note that a random clustering assignment would get a purity score from 0.25 to 0.3, since there are four roughly balanced labels in the IMDB data set. The trend for the three baselines is similar to the one of the CORA data set. From the results of both data sets, it is clear that {\em GSSClu} works especially well when a reasonable number of data points is received, because the optimization framework {\em DMO} can dynamically adjust the weights and compute a meaningful unified distance metric. In the meantime, {\em GSSClu} is superior to {\em GMicro} and {\em GSSClu [w/o opt.]} with the stream, and the sketch-based estimation is very close to the exact computation. In other words, the differences between the estimated values from {\em SGS(C)} and the exact values calculated by {\em Disk-based GSSClu} are extremely small and do not typically lead to quantitative clustering difference.

\subsection{Efficiency Results}

   \begin {figure}  \vspace{-0.4in}
      \begin{tabular}{cc}
        \includegraphics[width=1.6in]{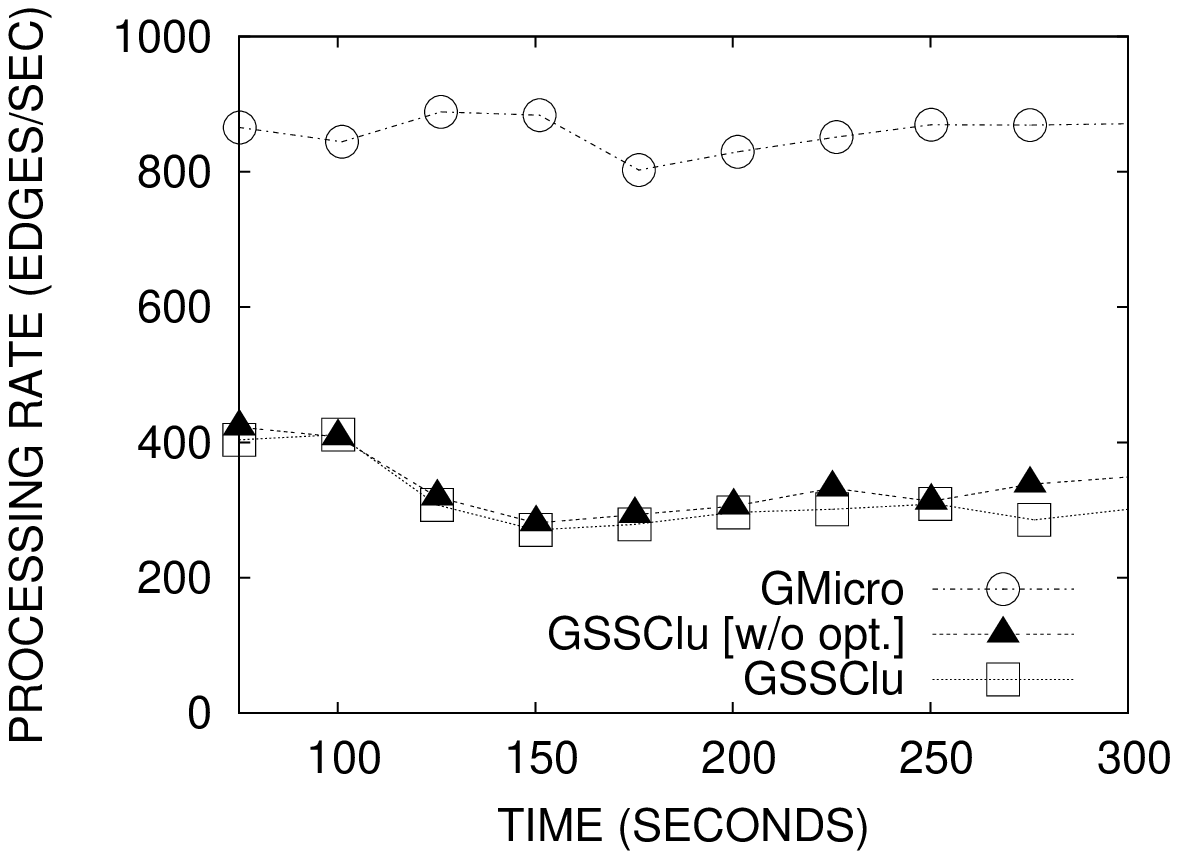} & \includegraphics[width=1.6in]{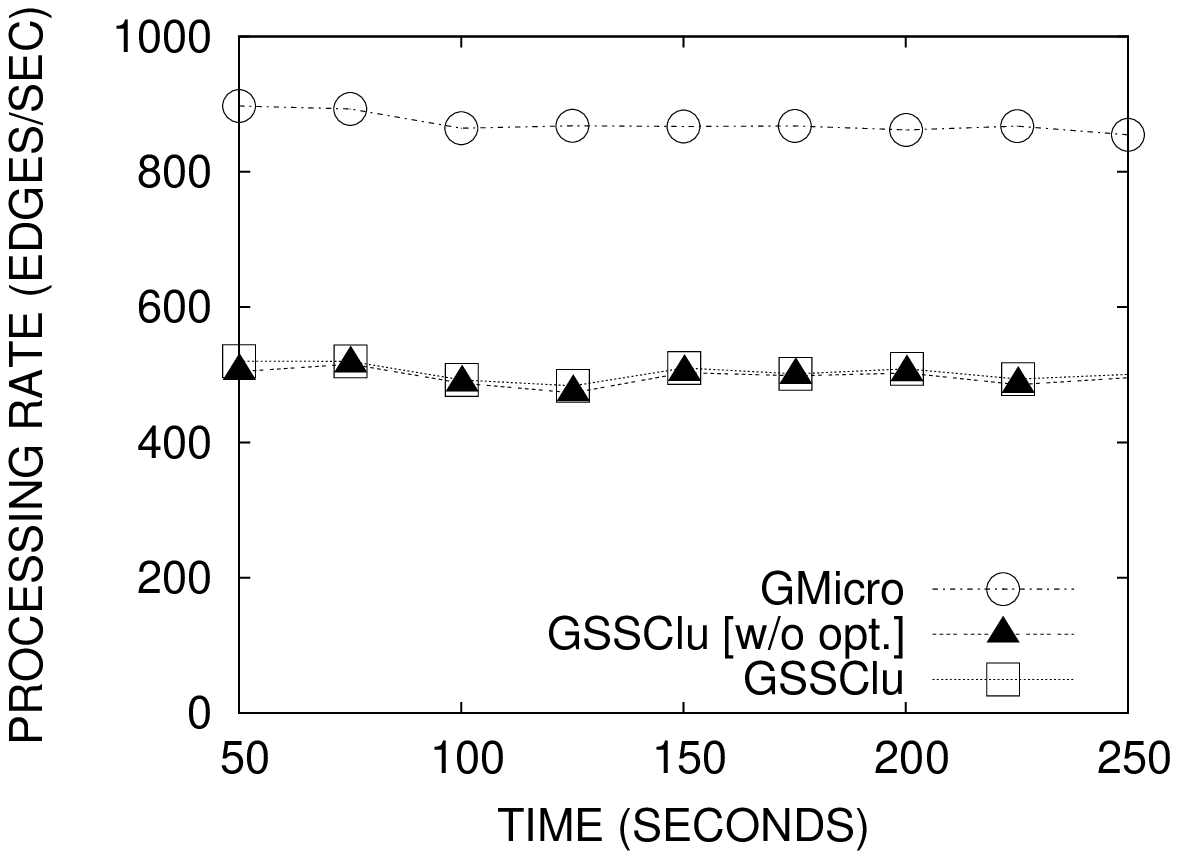}\\
      (a) CORA Data Set  & (b) IMDB Data Set\\
      \end{tabular}
    \caption{Efficiency Results}\label{time}
   \end {figure}
We also test the efficiency results of {\em GSSClu} and baselines on the real data sets. {\em Disk-based GSSClu} is 5 to 10 times lower than {\em GSSClu} due to the slow disk access. Thus, we do not show the efficiency result on {\em Disk-based GSSClu}. The results of {\em GSSClu} and other two baselines are shown in Figure~\ref{time} (a) and (b). In each figure, the X-axis shows the progression of the stream in terms of time, whereas the Y -axis illustrates the stream processing rate. The processing rate is computed based on the number of edges processed per second. The reason that we do not use the number of graphs processed per second is graphs could have skewed sizes. Some graphs are very large which need longer time to process, and then its low processing rate in terms of the number of graphs does not reasonably represent the underlying efficiency.

From both figures, one can observe that {\em GMicro} achieves the best efficiency. The reason that {\em GSSClu} based approaches consume more running time is {\em GMicro} only processes linkage data. The side information data has the same order of magnitude as the linkage data, which increases the running time of {\em GSSClu} based approaches. For example, in the CORA data set, each graph has 3.3 nodes in average, while it has 4.3 citations and 6.1 terms as the side information in average. In order to process such large number of additional side information, it is natural that the {\em GSSClu} based approaches consume more running time than the {\em GMicro} approach. Considering {\em GSSClu} based approaches, {\em GSSClu} and {\em GSSClu [w/o opt.]} process the same amount of data. It is evident that both {\em GSSClu} and {\em GSSClu [w/o opt.]} maintains a relatively stable processing rate with the progression of the stream. The figures show that the sketch-based statistics {\em SGS(C)} can indeed process stream data with high efficiency because the statistics remain the same memory consumption with constantly growing received graph objects. The low variability in processing rate is a clear advantage for use in practice. We further notice that {\em GSSClu [w/o opt.]} is slightly faster than {\em GSSClu}. This is quite natural, since {\em GSSClu} requires to periodically optimize and adjust the weights of links and side attributes. Since the optimization framework {\em DMO} is solved in the sketch representation, the optimization of {\em GSSClu} only adds a slight overhead on the running time. Considering the tremendous effectiveness improvement of the proposed approach {\em GSSClu}, the overhead of running time is quite acceptable.

\subsection{Sensitivity Analysis Results} \label{app_sen}

   \begin {figure}  \vspace{-0.4in}
      \begin{tabular}{cc}
        \includegraphics[width=1.6in]{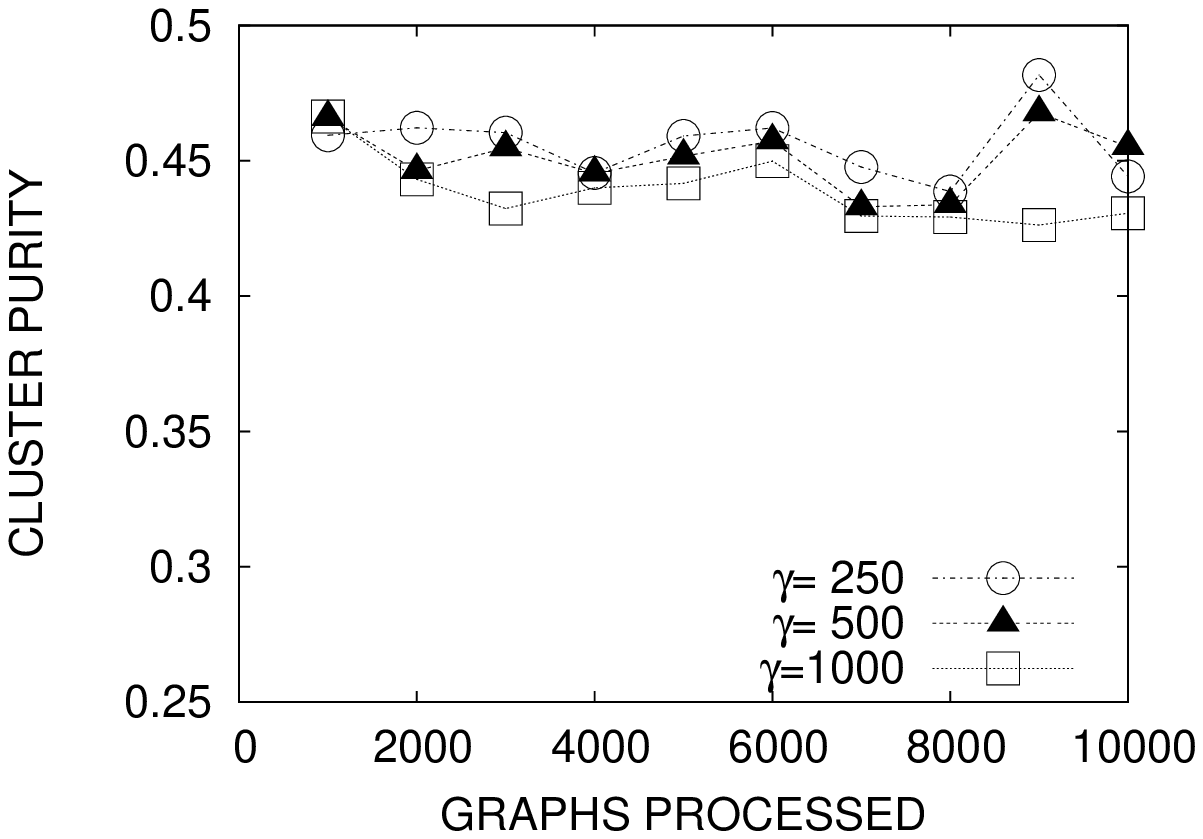} & \includegraphics[width=1.6in]{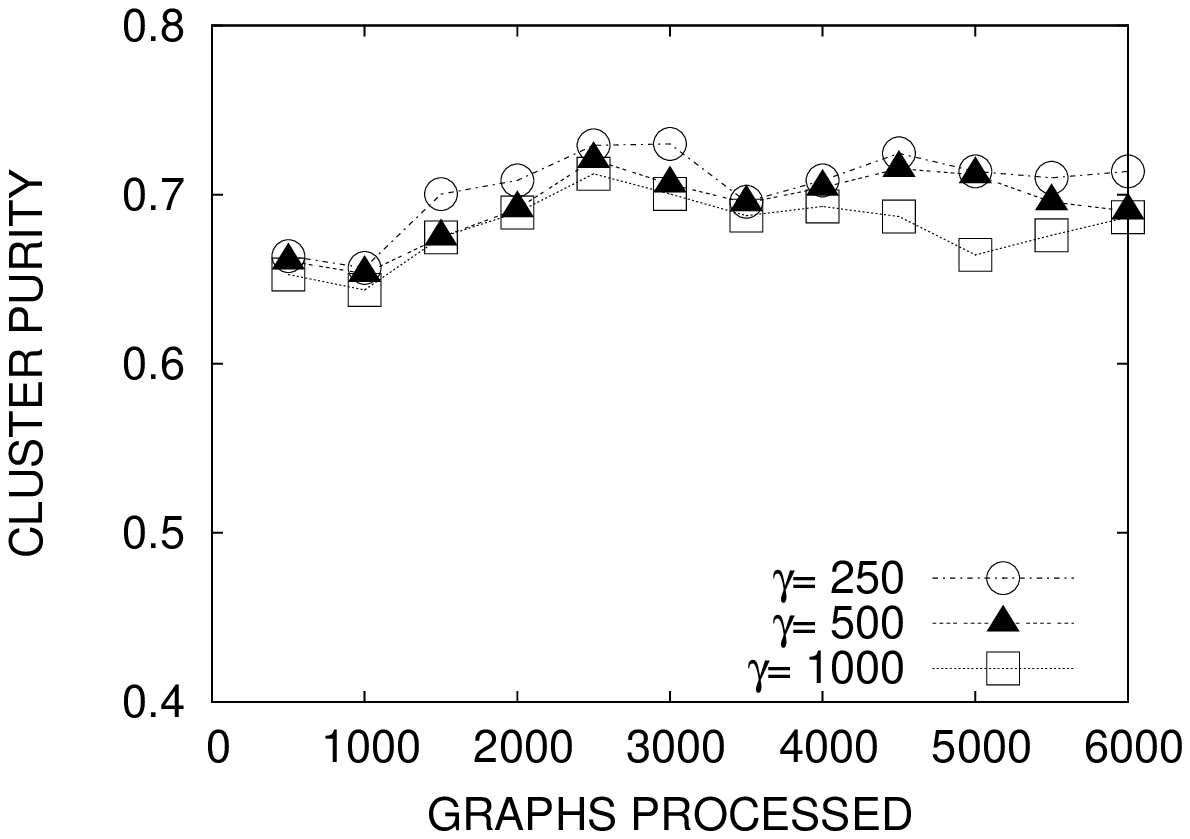} \\
      (a) CORA Data Set & (b) IMDB Data Set \\
      \end{tabular}
    \caption{Sensitivity Analysis with $\gamma$ on Purity}\label{gamma_purity}
   \end {figure}

In order to study how the parameter can affect the performance of {\em GSSClu}, we further conduct sensitivity analysis with respect to $\gamma$. As shown previously, $\gamma$ determines the frequency to update and adjust the weights in the E-S distance computation. Thus, it is reasonable to optimize the weights for every a few hundred received graphs. In Figure \ref{gamma_purity}, we present the effectiveness results on both data sets with three variations of $\gamma$ values, namely 250, 500 and 1000. The number of graphs processed is shown on the X-axis, and the cluster purity is illustrated on the Y-axis. In order to present the detailed differences among different settings of $\gamma$, the cluster purity is plotted on a more enlarged scale. From both figures, it is evident that {\em GSSClu} maintains stable quality for a wide range of parameter $\gamma$ settings. In the meanwhile, a smaller $\gamma$ value  slightly improves the cluster purity. The reason is that the weights and E-S distances can be adjusted more promptly under a smaller $\gamma$ setting to adapt the incoming data. All these suggest that our proposed method is not sensitive to the setting of $\gamma$ with respect to the effectiveness and {\em GSSClu} is quite robust with the progression of the stream.

We also test the efficiency of the proposed method over different settings of $\gamma$. Similar to Figure~\ref{time}, we plot the time in seconds on the X-axis, and the processing rate with respect to edges per second on the Y-axis in Figure~\ref{gamma_time}. We use a smaller granularity on the processing rate than that of Figure~\ref{time} to show the slight differences among three variations in $\gamma$. Based on the results shown on both figures, the processing rates for different settings of $\gamma$ are relatively stable with the increasing number of received graphs. Furthermore, it is evident that a smaller value of $\gamma$ can lead to lower processing rate of {\em GSSClu}. This is quite natural, since more frequent optimization consumes more running time. However, the overhead of the optimization framework {\em DMO} is minimal and does not affect the overall efficiency much. This suggests that the {\em GSSClu} approach is an efficient and scalable algorithm over a wide range of parameter $\gamma$ settings.

   \begin {figure}  \vspace{-0.4in}
      \begin{tabular}{cc}
        \includegraphics[width=1.6in]{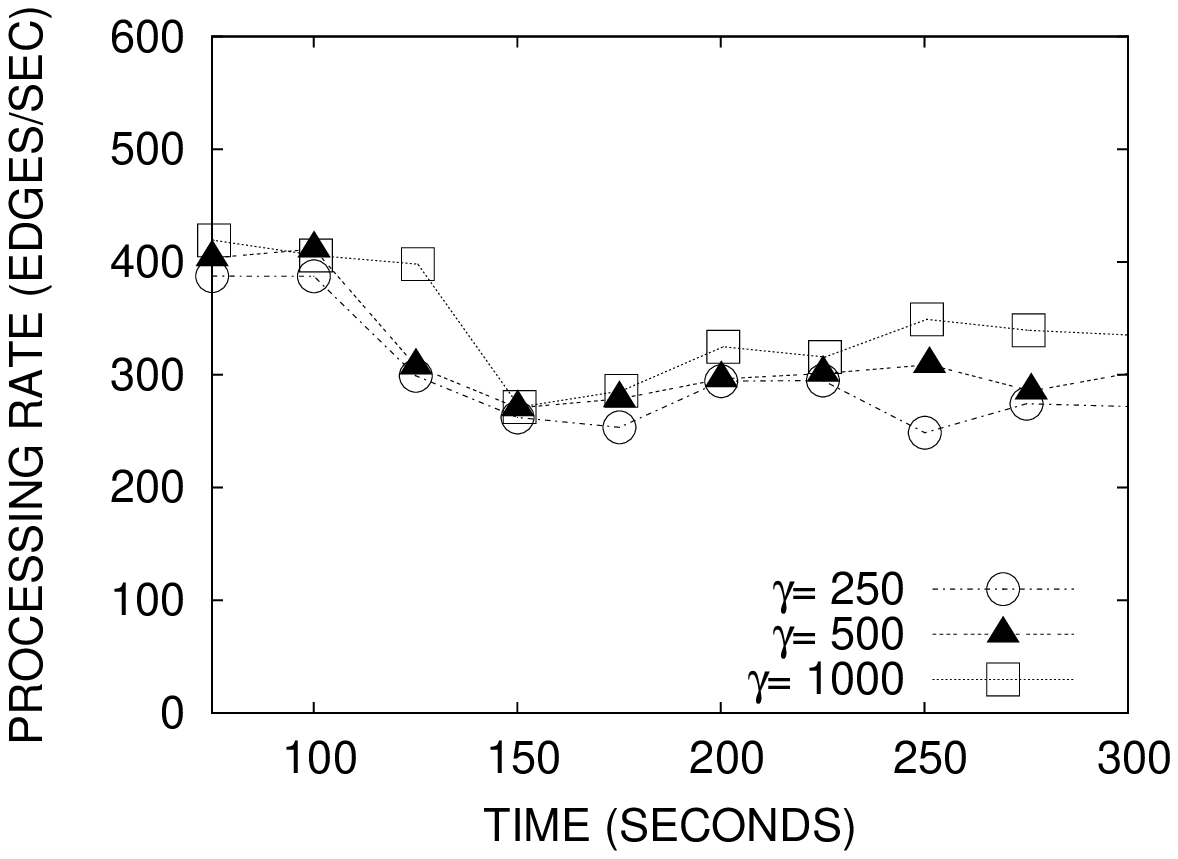} & \includegraphics[width=1.6in]{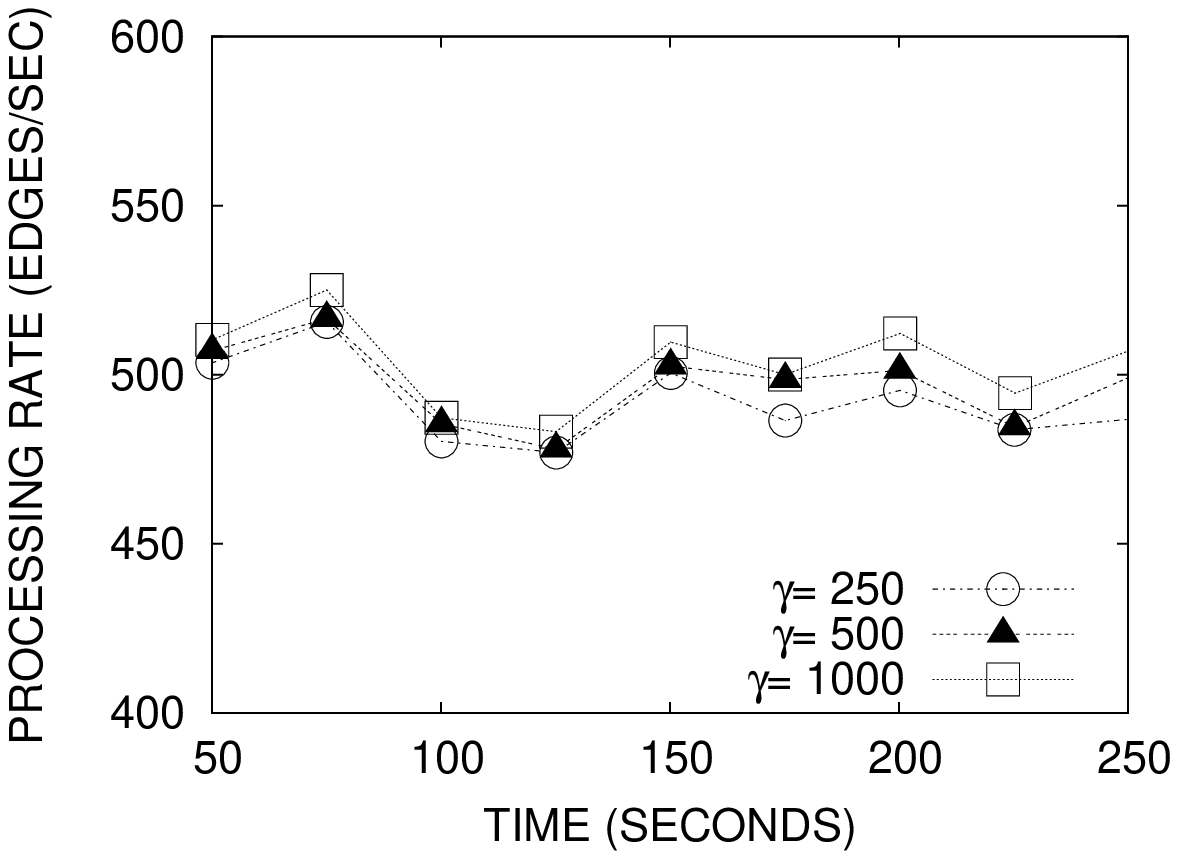} \\
      (a) CORA Data Set & (b) IMDB Data Set \\
      \end{tabular}
    \caption{Sensitivity Analysis with $\gamma$ on Efficiency}\label{gamma_time}
   \end {figure}

We further perform the sensitivity analysis with regard to the number of clusters $k$. We first show the effectiveness results with the number of clusters in Figure~\ref{cluster_num_purity}. We present the number of clusters $k$ on the X-axis, and the cluster purity score of the whole data set on the Y-axis. It is evident that the cluster purity score increases when we increase the number of clusters. The reason is that a larger number of clusters will generate clusters with finer granularity. In the meanwhile, we can observe that the cluster purity score increases only by about 0.03 even when the number of clusters is doubled. In other words, the cluster purity is highly consistent across all settings of $k$.  This suggests that the proposed approach {\em GSSClu} can constantly perform well under a variety of $k$ settings, and its effectiveness is not sensitive to the value of the number of clusters.

The efficiency results with the number of clusters $k$ are illustrated in Figure~\ref{cluster_num_time}. The number of clusters is shown on the X-axis, and the processing rate of the whole data set is presented on the Y-axis. We test the efficiencies with $k$ ranging from 10 to 18 for the CORA data set and from 6 to 14 for the IMDB data set. From the results on both data sets, it is clear that the {\em GSSClu} approach scales linearly with the number of clusters $k$ in terms of efficiency. Specifically, the smaller the number of clusters, the higher the processing rate achieved. This is because the distance computation in {\em GSSClu} scales linearly with the increasing number of clusters.

      \begin {figure}  \vspace{-0.2in}
      \begin{tabular}{cc}
        \includegraphics[width=1.6in]{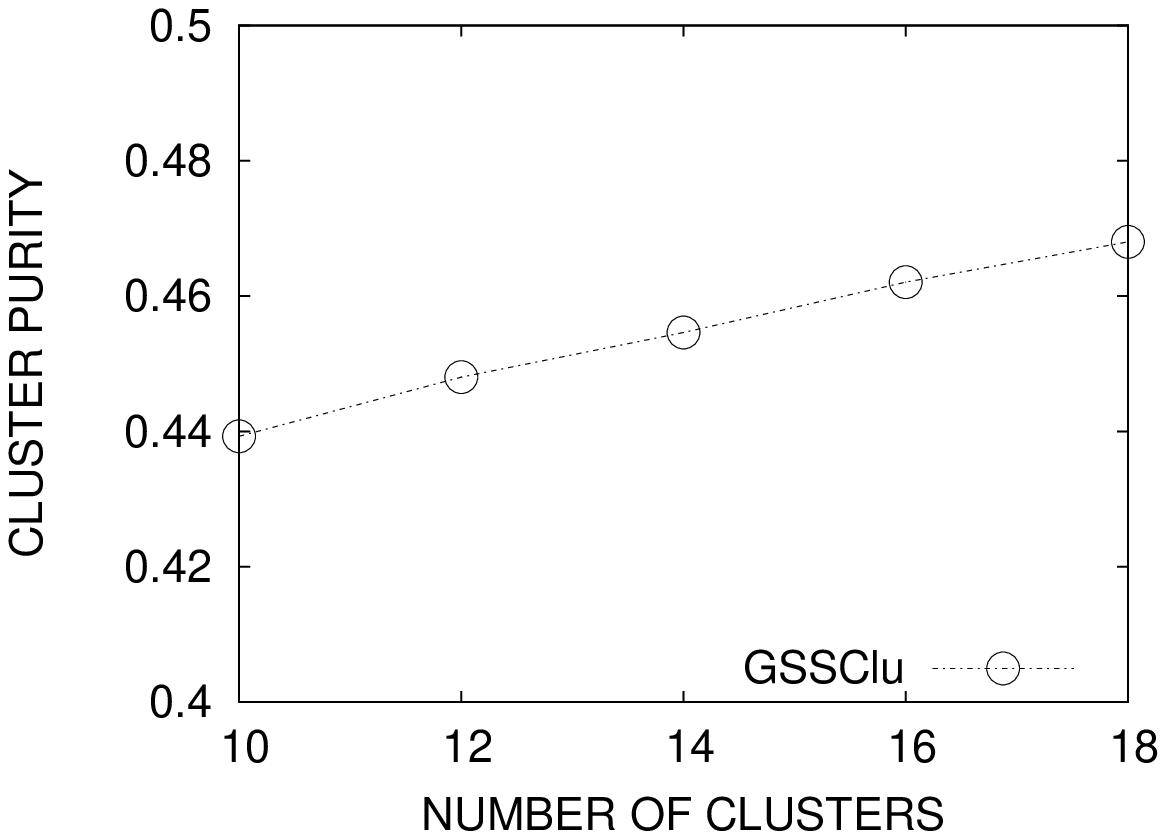} & \includegraphics[width=1.6in]{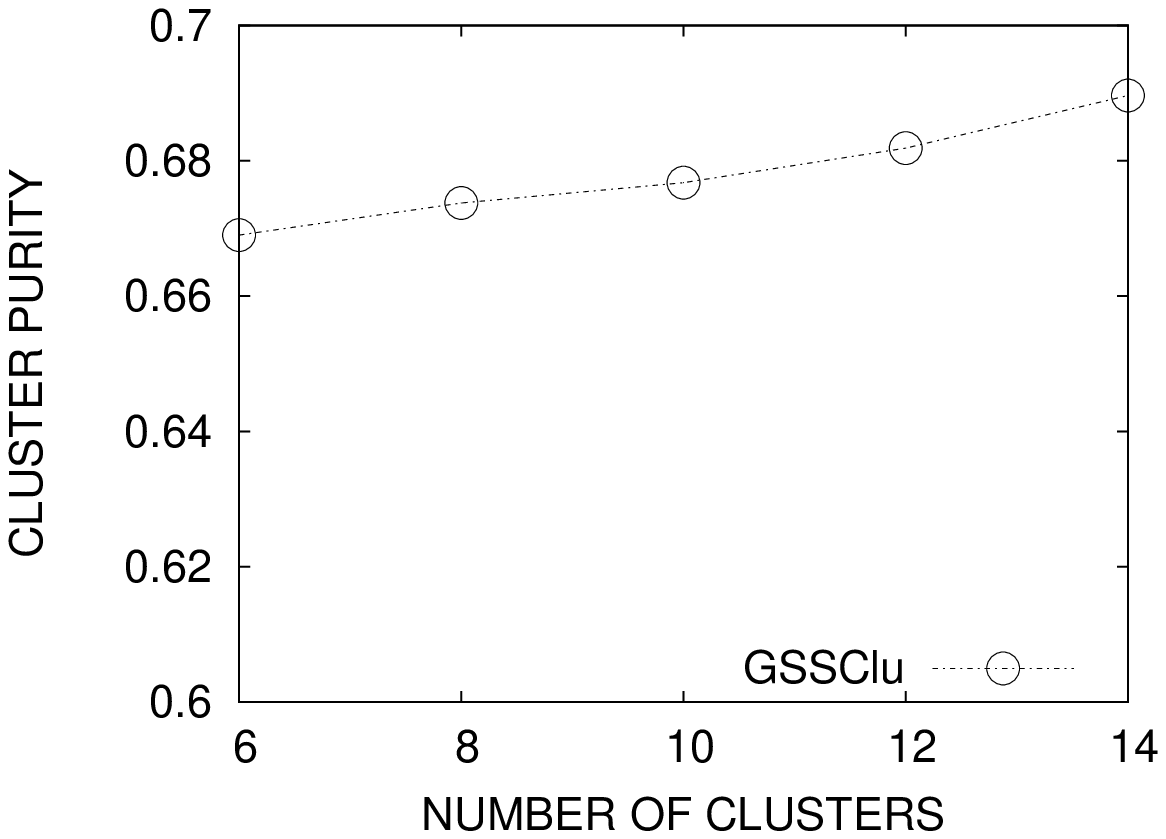} \\
      (a) CORA Data Set & (b) IMDB Data Set \\
      \end{tabular}
    \caption{Sensitivity Analysis with the Number of Clusters on Purity}\label{cluster_num_purity}
   \end {figure}

   \begin {figure}  \vspace{-0.2in}
      \begin{tabular}{cc}
        \includegraphics[width=1.6in]{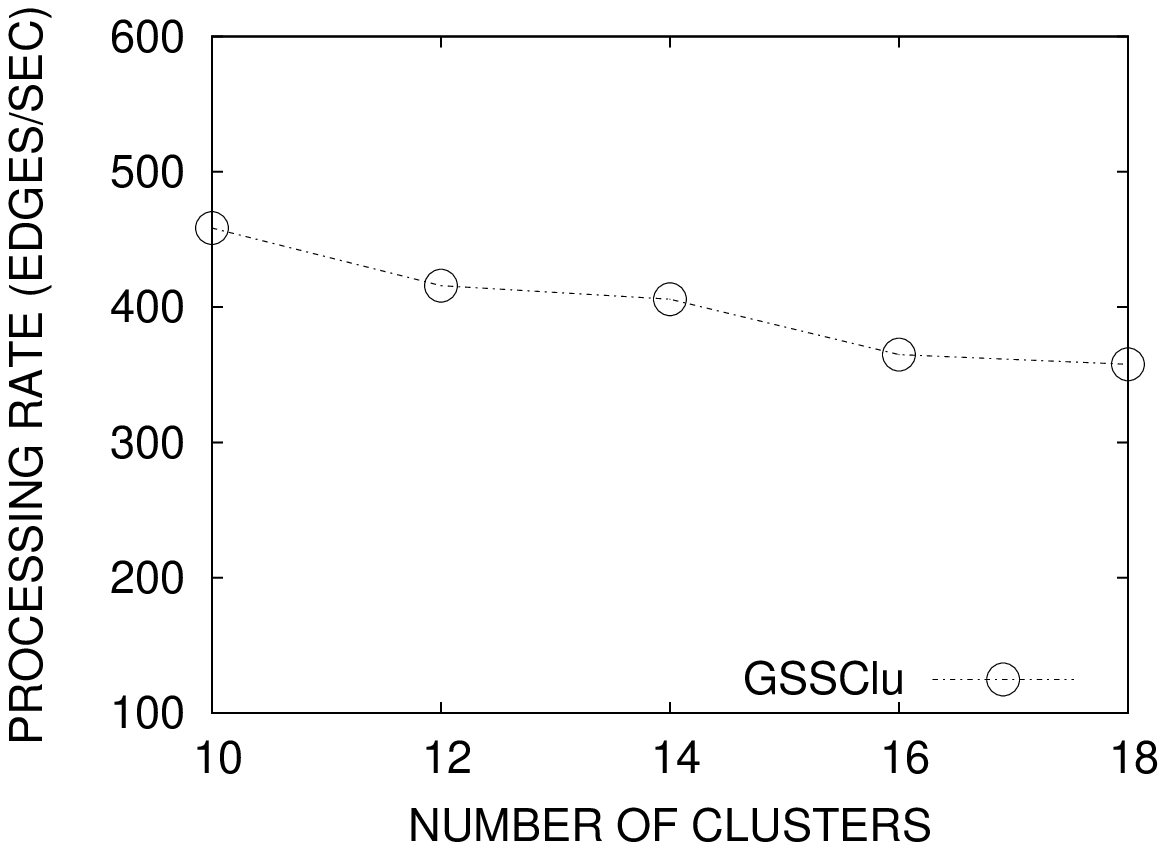} & \includegraphics[width=1.6in]{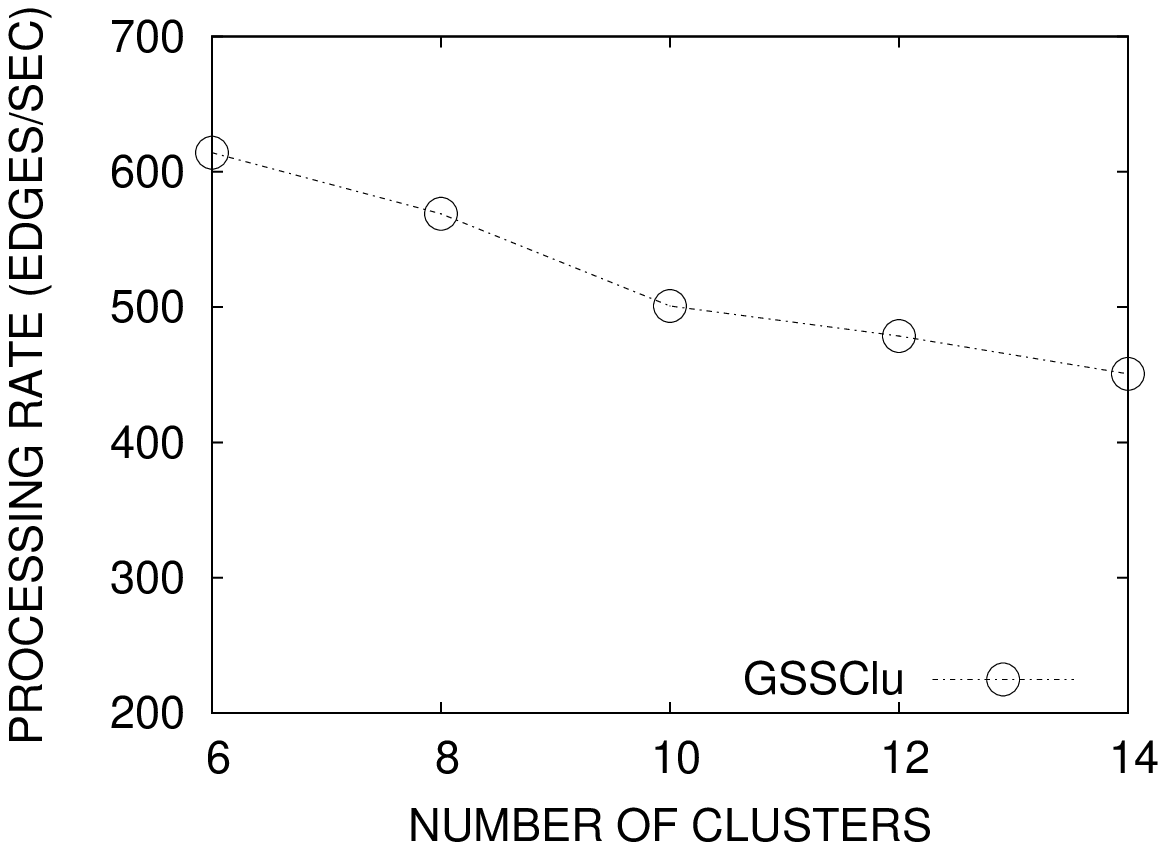} \\
      (a) CORA Data Set & (b) IMDB Data Set \\
      \end{tabular}
    \caption{Sensitivity Analysis with the Number of Clusters on Efficiency}\label{cluster_num_time}
   \end {figure}

\section{Conclusion} \label{conclusion}
In this paper, we present the first approach to cluster graph streams with side information. While many approaches have been devised to mine graph streams, they solely focus on the link structures of graphs. Many graph objects in real applications contain various forms of side information, which may be used to improve the clustering process. The problem is challenging, because not only it requires to process high volume links and side information with efficiency, but also it is non-trivial to incorporate side attributes to the graph clustering process.  In order to use both links and side attributes for the clustering model, we define a unified distance metric {\em E-S Distance} based on edges and side information. We further propose an optimization framework {\em DMO} to dynamically refine the distance metric by measuring the inter and intra cluster distances. A sketch-based framework $SGS(C)$ is also introduced to store the statistics of both edges and side information. We demonstrate that $SGS(C)$ can not only estimate the measures used in the clustering algorithm, but also solve the optimization framework {\em DMO} efficiently. The experiment results show that the proposed method significantly outperforms the baselines in terms of effectiveness, while it also maintains high efficiency and scalability. In our future work, we will consider using side information to improve other graph stream mining tasks, including classification, outlier detection and query processing.

\section{Acknowledgment}
This work is supported in part by NSF through grants IIS-0905215,
CNS-1115234, IIS-0914934, DBI-0960443, and OISE-1129076, US
Department of Army through grant W911NF-12-1-0066, Google Mobile 2014
Program and KAU grant.

\balance


\bibliographystyle{IEEEtran}
\bibliography{gcluster_side_sdm_long}

\end{document}